\documentclass[12pt]{article}
\usepackage{}
\newcommand{\abz}{\hspace*{.7cm}}
\author{Gerald Horwitz\thanks{Supported by BSF Grant 89-00244.} \\
University of Hannover, Germany\\
Racah Institute of Physics, The Hebrew University\thanks{Permanent Address.}\\
Jerusalem 91904, Israel\\
E-mail:geralda@vms.huji.ac.il}
\title{Time and Entropy from Semi-classical Tunneling  of the
Cosmological Scale Function\footnote{Lecture presented at the Moscow Quantum
Gravity Seminar,
Moscow, June 12-19,1995.} }
\begin{document}
\maketitle
\begin{abstract}
Two major apparently unrelated problems, that of the origin of time in
the universe associated with quantum gravity and of the entropy in de Sitter
 cosmological models,
are found to have their origin in a single physical phenomenon:
the semi-classical tunneling through a classically forbidden region of
the cosmological scale factor.  In this region there is a mixing of the
states of quantum matter and those of the semi-classical gravity which
 produces a thermal mixture of the matter states and hence an "entropy;"
this same mixing effect  brings about the conversion of a parametric
time variable into a physical intrinsic  time.
\end{abstract}
\section{Introduction}

\abz Black holes and certain cosmological models have
been under extensive and intensive study in recent years
in the context of Bekenstein-Hawking type thermal states, i.e.
thermal states associated with an event horizon.  These studies involve
general relativity (at the borderline with its quantum character),
quantum field theory and statistical mechanics.  There remain many
questions about these systems of a very fundamental character and though
most of the effects involved are not readily observable, they
nonetheless represent a laboratory for a great number of fundamental
questions in each of the separate disciplines as well as to the
connections between them.  What is the source of the thermal state in
these cases? Are they really thermal states: are these configurations
which look like thermal states really pure
quantum states or do they represent a density matrix? If they do how
does one reconcile an evolution from a pure state to a density matrix?
Is the level of semi-classical gravity sufficient to explain the effects?
For the cosmological case there is the additional question of starting
from a quantum gravity state of the wave function of the universe
for which there is no evidently no time
dependence and proceeding to the time evolution of the universe; this is called
in the
literature the problem of time.  There is, of course, here as well as in
general the problem of time direction. The number of papers dealing with both
the
subject of entropy source for black holes and related problems as well as the
efforts devoted to an answer to the "problem of time" are numerous and varied
and
to extensive to be included in the present work.\cite{time}

We wish to present here an approach which ties these problems to a
particular model of the universe and is carried out in a special
approximation. Thus we take an extreme position of particularity.
This has the disadvantage that it requires special conditions to hold, but they
do happen to be close to what corresponds to the real universe.  The second
advantage is
that even if it is not the correct explanation, the fact that one can out the
results
explicitly will add significantly to the understanding of more general and
models worked
out less completely.
We will argue that the justification of using semi-classical
 gravity is that it is necessary for time to be meaningful in this
 context, i.e. for an eigenstate for the whole universe.  The time seen in the
 universe, the cosmic time  may thus be explained without it necssarily being
the
 answer the question of time in quantum gravity in
general. That question may be quite independent of the subject of the time
evolution
in our observable universe.

We are able to give an answer to a considerable group of the above
questions in the context of a single physical idea,  the semi-classical
tunneling of the cosmological scale function out of a classical forbidden
domain.   In this domain quantum matter states are mixed with those of
semi-classical gravity yielding a thermal mixture of the  matter states
and  a further thermal contribution of the gravity associated with this
thermal mixing of the matter. Gravity in itself remains adequately described
by the single homogeneous function, the multiplicity comes from the
matter states. On the same basis gravity becomes semi-classical via the
mixing of highly excited levels of the matter states. The mixing of
matter states produces an effect that the parametric time represents a
reciprocal
temperature of the thermal mixture of the matter in the forbidden regime
and behaves like a physical time in the Lorenzian region when the scale
function penetrates the barrier.
The time  evolution follows the semi-classical orbits of the gravity and there
is a
foliation of the modified spacelike slices which reappear in the
semi-classical description.

Brief statement of results:
The various problems are solved on the basis of a semi-classical tunneling
occurring on small scales of the universe in going from a classically forbidden
 region to a classically permitted region of the cosmological scale function
  denoted $a(t)$. We are dealing with quantum matter in the form of
 conformally coupled scalar bosons, with mass and self-interaction.
  Gravity is taken to be that associated with a
homogeneous isotropic space-time (FRW metric) with a positive curvature
(k=1) and a positive cosmological constant $\Lambda(\Lambda_{eff})>0$ .
Though the sign of the cosmological constant is a necessary feature of our
theory, the sign of the curvature is one of convenience, as is the conformal
coupling of the bosons.  There then
exists a tunneling region for $a$  at a scale of the order of
magnitude of the de Sitter event horizon.  Our action
has the form\footnote{Units: $\hbar=c=G=1$}
\begin{eqnarray}
I=\frac{1}{16 \pi}\int_{}^{}d^{4}x \sqrt{|g|}[R-2\Lambda]+ \nonumber\\
\int_{}^{}d^{4}x\frac{\sqrt{|g|}}{2}\left[g^{\mu\nu}\varphi_{,\mu}\varphi_{,\nu}-\frac{R}{6}\varphi^{2}-
m^{2}\varphi^{2}+\lambda\varphi^{4}\right].
\end{eqnarray}
The metric, using conformal time is:
\begin{equation}
	ds^{2}=a^{2}(\eta)[d\eta^{2}-d\chi^{2}-sin^{2}\chi(d\theta^{2}+sin^{2}\theta
d\phi^{2})]
	.
\end{equation}

We study solutions of the
Wheeler DeWitt equation  in the above minisuperspace model.
Let us consider this as a realistic model of the universe. In past
work\cite{HW} and
\cite{BHW}
we have taken $\Lambda$ as a phenomenological constant.  Here we wish to
consider  a $\Lambda_{eff}$ which has its origin in a special solution of a
classical condensate for the $\phi$ which can occur for a Higgs mass
($m^{2}<0$). A phase transition which removes the condensate removes the
cosmological
constant. The entropy grows as a result of this transition. This work is in
progress.

We make a conformal transformation for the $\phi$ fields.
\begin{equation}
	\varphi= \phi/a  \qquad      g_{\mu\nu} = a^{2}\gamma_{\mu\nu} ,
\end{equation}  $\gamma_{\mu\nu}$ corresponds to the static Einstein metric.
Using the FRW solution we can write our action in the form:

\begin{eqnarray}
I=\frac{3}{16\pi}\int_{}^{}d^4x\sqrt{|\gamma|}[\dot{a}^{2}-a^{2}+\frac{\Lambda}{3}a^{4}]+
	\nonumber\\
\int_{}^{}d^{4}x\sqrt{|\gamma|}\frac{1}{2}[\gamma^{\mu\nu}\phi_{,\mu}\phi_{,\nu}
	-\phi^{2}-\mu^{2}a^{2}\phi^{2}+\lambda\phi^{4}].
\end{eqnarray}

We follow  Banks' \cite{Banks} analysis of the first order
Born-Oppenheimer  approximation, regarding
the semi-classical gravity as the heavy degree of freedom and the quantum
bosonic matter as the light degree of freedom, giving an effective Schroedinger
equation in terms of the parametric time variable obtained from solving
the equations of motion of the semi-classical gravity. This time is shown to be
associated with a foliation of modified spacelike hypersurfaces.
It is essential to note that the modified Schroedinger equation
obtained here is dependent on the mixing of the semi-classical gravity
 with the quantum matter degrees of freedom!

 One universe, considered as a unique one which represents a closed system,
 being in an eigenstate has \underline{no time} regardless of
 whether the eigenvalue is zero, i.e. this holds quite separately of the
arguments about
the constraint conditions.  Notice also that even if the eigenvalue
 is zero, this does not imply that the operator is zero, there may be many
 states corresponding to this eigenvalue. Such state can only acquire a
 time by some of the states going semi-classical and this can only occur when
 some sum over highly excited states of the degree of freedom which is to
become
classical, or as here
when other states are mixed with this degree of freedom.
 It is unphysical for the state of the universe at time scales a few orders of
magnitude
greater than Planck time to treat the matter as well as gravity
semi-classically
unless there is some condensate of the matter.  The only feasible time variable
for
this situation is an intrinsic time variable of the kind that we obtain.

 We have approached the problem in three different complementary ways.
 Each provides insight and understanding of aspects of the problem less obvious
 in the other approaches.  We have made a statistical mechanical evaluation
 of the entropy\cite{HW}, thus showing that what we obtain is a real
 statistical state.  We have analyzed the problem in terms of the
 wave function of the universe approach by two different ways,
Born-Oppenheimer
 analysis\cite{BHW} of the wave function of the universe and the
 evaluation by a functional integral approach.  The Born-Oppenheimer
 approach with the inclusion of the leading higher order term leads to
 the understanding that the time variable which is the solution of the
 classical equations of motion is also a time variable for the
 Schroedinger equation for the matter.  This holds, however, only if the
 matter wave function is not an eigenstate of the matter Hamiltonian.
 Thus the conversion of the parametric time to the time of the
 Schroedinger equation requires a mixing of matter states, here by the
 intermediary of mixture with the semi-classical gravity.  This feature is
 not at all obvious in the functional integral approach, and, in
 fact, has been overlooked by many people using the semi-classical gravity
 as a source for a time variable. We evaluate the probability weight of being
at
a particular scale function $a$;this quantity we denote $W(a)$.
This is presumed to describe the state of the universe at cosmic scales some
orders
of magnitude greater than Planck lengths. We evaluate a quantity which is
formally a density matrix.
That quantity may be equivalent to the scalar product of the wave function of
the universe
with itself if certain fluctuations are small enough. Similarly there is
another step in
the procedure which involves interchange of the order of integration between a
time
variable fixing the constraint and an integral over matter configurations. In
terms of the
quantity calculated we find a well defined intrinsic time variable and a
thermal admixture of
the matter the above calculated quantity being the entropy when $a$ has a value
greater
than or equal to the outer turning point of the cosmological scale function.

In section 2 we will review some of the special properties of
gravitational statistical thermodynamics as a background for our specific
discussion . This aspect of the problem is the most
 carelessly treated feature in the literature of this subject.
 In Section 3 we consider the question of an appropriate
energy and Hamiltonian for our model and then write down an expression
for the corresponding entropy. Section 4 introduces the Wheeler-DeWitt equation
and the constraints; we introduce the concept of the semi-classical, intrinsic
 foliation to replace the extrinsic one which has been eliminated by the
diffeomorphism
invariance. Section 5 deals with a higher order
Born-Oppenheimer approximation in which a parametric time is to be
identified as physical time; this parametric time also has a possible
imaginary domain where it will subsequently associated with a reciprocal
temperature. The functional integral evaluation of $W(a)$ is presented in
section 6, where the
contribution from the tunneling region is identified with the exponential
of the entropy. This quantity is analyzed in terms of the relation between
density
matrix and a pure state. Conditions are discussed under which this could be in
effect
equivalent to a pure state; fluctuations of the wave function at the turning
points of the
scale function must be sufficiently small.
The last section concludes with a summary and a discussion of further
work and various problems about the above work.
\section{Gravitational statistical thermodynamics}

\abz Is there any thermodynamic equilibrium for gravitating systems? The
answer, we suggest, is a qualified yes. Namely, although there is no absolute
maximum of entropy valid for gravitational systems since its Hamiltonian
and action are indefinite quantities, there can be a local
maximum of the entropy which is sufficiently long lived and to which the
corrections are sufficiently small to be a useful description and
calculational device. Energy is always problematic quantity
 in general relativity. We would like to suggest that systems and geometries
 which lack some conserved energy-like quantity
cannot be expected to have any  thermodynamic
equilibrium configuration. Such a conserved "energy"
itself implies some special time and related temperature variable. In the
next section we will review a procedure to define such a conserved energy
for metrics that correspond to cases with a conformal time-like killing vector.
There must be some kind of generalization of the conserved energy being related
to
time translation invariance.

While many fundamental questions are asked of the quantum mechanical and
general
relativistic behavior of these systems,the treatment of statistical
mechanics is done very uncritically.
Gravitating systems cannot strictly be divided into subsystems in the
sense commonly used in thermodyamics to define canonical or grand
canonical ensembles.  If one is sufficiently far away from a phase
transition such division is not problematic. However, in terms of phase
transitions the open systems are not independent thermodynamic systems
with fixed intensive parameters given by a reservoir (e.g. temperature).
 It is very common that people doing the statistical
mechanics of black holes or cosmology begin with an expression for the
free energy as a representation of thermal behavior. There is
\underline{no} justification in taking the partition function as the
representative function of thermodynamic equilibrium for these systems.
This is commonly done by invoking conventional statistical
mechanics- separation of system into large and small subsystem with the
large subsytem being regarded as a heat bath and the small one as the
systems which has a canonical ensemble distribution. This separation is
fundamentally incorrect and the stability of the small subsystem is, in
general, not that of the positive heat capacity.
 For example, York's \cite{York} correction of Hawking's
canonical ensemble treatment of the stability of a black hole in a
radiation cavity is evidently the correct way to evaluate the canonical
ensemble where a temperature is fixed at the boundary surface at finite radius.
 This result appears to be independent of the universe external to the box, but
is it really independent? It is unclear under what circumstances
the York calculation could  be realized physically.
 An exception may be the case of a black hole
produced spontaneously by a density fluctuation.  In such case the
particle horizon  limits the region which could be effected by the black
hole and there the temperature is indeed fixed at the surface of the
particle horizon, producing a finite region of space inside of which the
black hole could be found in a stable canonical ensemble.
Hawking's result\cite{Haw} is correct for the case where the black hole and
its surrounding cavity are alone in the universe with the
temperature fixed at infinity.

Even in standard systems, it is the microcanonical ensemble which is
fundamental and is based on conserved macroscopic quantities, especially
the energy. Here the quasi-ergodic hypothesis serves as the fundamental
assumption and there has been established some reasonable justification
of this principle.  Other ensembles are derivative quantities and their
validation is based on the applicability of the thermodynamic limit for
conventional systems. Due to the long range attractive forces, this is
not a valid principle for gravity and so one must proceed with caution in
applying any conventional wisdom to gravitational systems.  This caution
is singularly lacking in many of the works written in this field.
 Just one example will be given at this point. If the starting point is
 the microcanonical ensemble one can derive the free energy, in a saddlepoint
 evaluation of the energy constraint.  If one begins from the canonical
 ensemble one can evaluate the entropy from a saddle point evaluation of
 the  partition function.   The common equation relating the two
is that which is valid in
 the thermodynamic limit, namely the Legendre transformation.   However,
 the two ensembles being  inequivalent for gravitational systems, the
 fluctuations differ and the two different ways of evaluating the entropy
 are not equivalent!

\section{Energy and entropy of the cosmological \\model}
\abz Horwitz and Katz\cite{HK} found a conserved energy for FRW cosmological
metrics
using a reference metric scheme developed earlier by Katz, Lynden-Bell
and Israel\cite{KLI}. The key to this approach to defining a conserved energy
is to
introduce a reference metric which has a static time-like Killing vector.
If the physical metric can be mapped smoothly onto the reference metric
without singularities, then the conserved current defined relative to
the reference metric can be shown to lead to a conserved energy which
corresponds to the physical energy for all case where an energy is known.
Katz etal \cite{KLI} dealt with the case of asymptotically flat space-times.
The conserved
energy is a function of the metric only.  For the cosmological case, at
least for a closed universe, this constant is shown to be equal to zero.
A covariant Hamiltonian was derived which was shown to satisfy the
following identity:

\begin{equation}
	H=E+\int_{}^{}\sqrt{|g|}(T^{\mu}_{\nu}-G^{\mu}_{\nu})\xi^{\nu}_{conf}
d\Sigma_{\mu}
	=D,
\end{equation}
where the symbol D represents the dilatation generalized to include
gravity as in the work of \cite{HW}; the energy being zero, the Hamiltonian
can be identified with the dilatation.
The Hamiltonian is derived from the action
\begin{equation}
	I=\frac{1}{16\pi}\int_{}^{}(R-2\Lambda)\sqrt{|g|}d^{4}x+\int_{}^{}d^{4}x
	[\partial{_\alpha}(k^{\alpha}\sqrt{|g|})]+I_{matter}.
\end{equation}
This vector \(k^{\alpha}\), which depends on both metrics in the action
produces the same effect as the York-Hawking surface term to eliminate
the second derivatives except that here the reference space appears explicitly.
The formalism used is to express all quantities in terms of the reference
metric like the approach developed by Rosen\cite{RO} for a different purpose.
The modified Christoffel symbol \(\Delta^{.}_{..}\) is defined (the bar
referring to
the reference metric)
\begin{equation}
	\Delta^{\mu}_{\nu\lambda}=\Gamma^{\mu}_{\nu\lambda}-\bar{\Gamma}^{\alpha}_{\mu
\nu }
\end{equation}
and then
\begin{equation}
	k^{\alpha}=g^{\alpha\beta}\Delta^{\nu}_{\nu\beta}-
g^{\mu\nu}\Delta^{\alpha}_{\mu\nu}.
\end{equation}

If we now quantize semi-classically; our identity remains, with the conserved
energy
depending only on the metric remaining zero and:
\begin{equation}
	\hat{D}_{ren}=\hat{H}_{ren},
\end{equation}
 and the renormalized Hamiltonian remains equal to the conserved energy E=0
in the sense that

\begin{equation}
	\hat{H}|\Psi>= E|\Psi>=0.
\end{equation}

We can define the entropy by an assumed quasi-ergodic theorem, that the
entropy is the log of the sum over states satisfying the constraint that
$D_{ren}=0$.
Thus we have
	\begin{equation}
			expS= Tr(\delta(\hat{H}-E))=Tr(\delta(\hat{D}))
	\end{equation}
for our case,where we have restored the E which may differ from zero
for example, for nonclosed universes; our entropy expression in terms of D
remains valid.

Horwitz and Weil\cite{HW} evaluated the cosmological entropy this way some
years ago
establishing a first principles calculation of the entropy. The result found
was
a generalization of the Gibbons-Hawking de Sitter entropy which was found to
 be thermodynamically unstable. The stable entropy is associated with a state
where
the universe has a finite density of thermal bosons. This increases the entropy
by
a modest amount.  Furthermore establishing the principle that  the quantity
being calculated is a thermodynamic entropy and not something that just
formally
looks like an entropy.

\section{Wheeler-DeWitt Equation}
\abz Writing the Wheeler-DeWitt equation  for our minisuperspace model
\begin{equation}
	(\hat{H}_{G}+\hat{H}_{M})\Psi(a,\phi)=0,
	\label{12}
\end{equation}
The Hamiltonian density\footnote{Properly for the Hamiltonian we
should replace the time derivatives by the corresponding momentum variables;
that connection is sufficiently obvious that we have chosen to retain
 this form for conciseness.}
 of the gravity is
\begin{equation}
	h_{G}=-\frac{3}{16\pi}(\dot{a}^2+V(a))
\end{equation}
where the gravitational potential is
\begin{equation}
	V(a)=[a^{2}-\frac{\Lambda}{3}a^{4}].
\end{equation}
The matter Hamiltonian density is here
\begin{equation}
	h_{M}=\frac{1}{2}(\dot{\phi}^{2}+(\nabla \phi)^{2})+
	\mu^{2}a^{2}\phi^{2}-\Lambda\phi^{4}).
\end{equation}

The Wheeler-DeWitt approach begins by making a 3+1 split of the spacetime and
considering the Hamiltonian defined on an arbitrary spacelike hypersurface.
Setting up a foliation of the
space-like hypersurfaces, can be considered to define the time variable.  In a
minisuperspace model a preferred foliation might  be associated with a
time-like Killing vector. The Hamiltonian in arbitrary coordinates can be
expressed in terms of a lapse function associated with a normal to
the spacelike hypersurface and a shift vector associated to transformations
on the hypersurface. In general using considerations of symmetry
one can make arbitrary change of the lapse function  and arbitrary
coordinate transformations on the space-like hypersurface.  These
invariance properties lead to 4 constraint equations denoted respectively
as the Hamiltonian and momentum constraints.  For general relativity the
Hamiltonian can be written completely in terms of the constraints and the
lapse function and shift 3-vector. We can quantize by usual
canonical methods,  finding the momenta and replacing the classical
momenta by the derivatives with respect to canonical coordinates:
\begin{displaymath}
	p_{a}\sim \dot{a}\longrightarrow \frac{\delta}{\delta a}
\end{displaymath}
\begin{displaymath}
p_{\phi}\sim\dot{\phi}\longrightarrow \frac{\delta}{\delta \phi},
\end{displaymath}
One commonly follows the Dirac method and associates the constraint with
a zero eigenvalue of the constraint operators.
The constraint equations and the wave function of the universe are functions of
the
induced metric $h_{i,j}$ on the spacelike hypersurface and the matter
coordinates. We proceed with the Dirac approach where we transfer the
constraint condition to an eigenvalue problem on constraining our problem
to be the zero eigenvalue of the  Hamiltonian and momentum constraint
operators acting on the wave function. The  Hamiltonian, consisting of a
gravity and a matter part when acting the wave function are familiarly
known as the Wheeler-DeWitt equation:
\begin{equation}
	[\hat{H}_{G}+\hat{H}_{M}]\Psi(h_{i,j},\phi^{\alpha})=0
\end{equation}
where the $\phi^{\alpha}$ are here our boson matter fields.
Corresponding equations exists for the momentum constraints.

What happens to space-like hypersurfaces when we apply the constraint
condition? One might consider two possibilities: either the space-like
hypersurfaces remain distinct and the quantization which is carried out on
 some arbitrary one to is retained with transfer operator to any
other surface absent. Alternatively we might consider that the wave function
satisfying
the constraint equation includes contributions from all the hypersurfaces.
 It is easy in the functional integral
formalism below of (\ref{FI}) to see that the latter is the case. The result
for the fixed energy case is a superposition of the multitude of spacelike
hypersurfaces. We will return to this explicitly below after we have presented
the
functional integral evaluation below.
This is manifestly a superposition of the amplitudes summed over the
original leaves. Choosing a common saddlepoint of $a(t)$  and $t$
 in the functional integral, we precisely get a path through the modified
hypersurface evaluated at $a$'s  determined by the classical equations. The
interchange of orders of integration when justified as in our case,
carries us over from a many-fingered time situation which has not such
simple interpretation to a well defined time variable and a well defined
path once we have specified initial value conditions and assuming that
there is a unique choice, or a well-determined basis to choose one of
several possible choices.
\section{Born-Oppenheimer analysis}

\abz The essential approximation of Born-Oppenheimer is to assume that
the wave function can be written as a product wave function of matter and
gravity parts.  The zeroth approximation has the matter wave function
depending on the gravity part only parametrically.  We shall be
interested in including the next order approximation.

\begin{equation}
	\Psi(a,\phi)=\chi(a,\phi) \psi(a).
\end{equation}
Then the product wave functions are determined by the 1st order
Born-Oppenheimer equations:

\begin{equation}
\frac{\left(\chi|\hat{h}|\Psi\right)}{\left(\chi|\chi\right)}=(\dot{a})^2+V(a)
+\left\langle\hat{h}_{M}\right\rangle=0.
	\label{EC}
\end{equation}
where
\begin{equation}
	\left\langle\hat{h}_{M}\right\rangle=
\frac{\left(\chi|\hat{h}_{M}|\chi\right)}{\left(\chi|\chi\right)}.
\end{equation}
The second equation is then found to be
\begin{equation}
	\frac{\left(\psi|\hat{H}|\Psi\right)}{\left(\psi|\psi\right)}=
	\left[\hat{H}_{M}-\left\langle\hat{H}_{M}\right\rangle\right]\chi(a,\phi)-
	\frac{\partial{(ln\psi)}}{\partial{a}}\frac{\partial{\chi}}{\partial{a}}=0.
	\label{A1}
\end{equation}
Since we are dealing with semi-classical gravity then we have
\begin{equation}
	\frac{\partial{(ln \psi)}}{\partial{a}}=i\dot{a} \qquad
	\frac{\partial{(ln \psi)}}{\partial{a}}\frac{\partial}{\partial{t(a)}}=
	\frac{\partial{}}{\partial{t(a)}}
	\label{A2}
\end{equation}
where the quantity $t(a)$ is determined from the solution of (\ref{EC}) and is
to
be identified as a parametric time variable.  It can be either real or
imaginery.  We choose to identify it with the physical time of the
universe on the basis of (\ref{A1})-(\ref{A2}) being in effect a Schroedinger
equation for
the matter wave function. Notice, however, one feature of the result,
namely that there is no Schroedinger equation if the $\chi's$ are
eigenfunctions of the matter wave function.  The last term of (\ref{A1}) is by
its very nature a mixing term; in this case it mixes semi-classical
degrees of freedom with matter degrees of freedom.  Consequently our pure
wave function appears to have a superposition of all matter eigenstates.
As we shall see below this mixture is produced in the tunneling region
and the imaginery time of that region will appear as a reciprocal temperature.
The mixture of the matter states will be a thermal admixture.   Solving
(\ref{EC}) for $t(a)$ we find the equation
\begin{equation}
t(a)=2\int_{a_{ref}}^{a}\frac{da}{\sqrt{-\frac{16\pi}{3}
\left\langle\hat{h}_{M}\right\rangle-
a^{2}+\frac{\Lambda}{3}a^{4}}}=
2\int_{}^{}\frac{da}{\sqrt{A(a)}}
\end{equation}
Then for $A(a)>0$, $t(a)$ is real, while for $A(a)<0$, $t(a)$ is imaginery;
$A=0$ gives the turning points.
\abz {Functional integral evaluation }

The Born-Oppenheimer approach was important in terms of emphasizing
certain physical features of the approximation used.  However, functional
integral evaluation is preferrable for other aspects of the problem.
The quantity which has the desired properties of yielding a semi-classical
thermal universe evolving with the time given by the Born-Oppenheimer
Schroedinger
eqution has the form of a density matrix which we will evaluate by functional
integral methods. This density matrix is potentially equivalent  to the square
of
the Wheeler-DeWitt wave function under conditions which we will describe, but
not
verify in the present paper. Thus we consider the weight function for the
universe to
be at a  scale $a$:
\begin {equation}
W(a)=\int_{}^{}D[\phi(\vec{x})]\int\limits_{}^{}\frac{d\tau}{2
\pi}\int\limits_{\scriptstyle
{\      a(0)=a(\tau ) \atop \enspace=a}}^{\scriptstyle {\  WKB}}
{D[a]}^{}\int\limits_{\scriptstyle
{\      \phi(0,\vec{x})=\phi(\tau,\vec{x})\atop \enspace=\phi(\vec{x})}}
D[\phi(x)]e^{-iI(\{a\},\{\phi(\vec{x})\})}.	\label{FI}
\end{equation}

The energy constraint is fixed by integration  over a ``time-like''
coordinate $\tau$. The semi-classical path of integration of $a$ proceeds from
the value of the argument of $W(a)$ to a turning point and back.
The choice of the reference turning point which corresponds to a  choice of
boundary
conditions and sets a direction to time is to choose the  lower turning point
in the
tunneling regin denoted  $a_{-}$.
However, to carry out our evaluation of this integral we proceed to one
further approximation.  We interchange the order of the $\phi$ integral
with that of the $\tau$ integral. Evaluating the time integral by saddlepoint
method,
we would otherwise have our time saddlepoint depend on $\phi(\vec{x})$.
Interchanging the orders allows us to convert the result from
a many fingered time  solution to a single time.  The justification for
this interchange has been discussed at some length by  Brout and
Venturi\cite{BV}.
The essential point is that the many degrees of freedom of the matter
must produce sufficiently small fluctuations to justify this step.
 We shall assume that conditions
which Brout and Venturi have found for its validity hold here.
Since both the $a$ and the $\phi$ are
periodic functions of $\tau$, if we carry out the following integral
within the tunneling region. Then
\begin{equation}
\int\limits_{\scriptstyle
{\      \phi(0,\vec{x})=\phi(\tau,\vec{x})\atop
\enspace=\phi(\vec{x})}}^{}D[\phi] e^{-I_{M}(\{\phi\},a)}
=e^{-\tau F}.
\end{equation}
This is the functional integral over the boson degrees of freedom for an
\(a\) inside the tunneling region giving us a partition function with the
inverse temperature equal to the return path in time from $a$ to $a_{-}$.
Thus the above expression for an $a$ in the tunneling region, with the $\phi$
integral carried out takes the form
\begin{equation}
W(a)=\int_{\tau_{0}-i\infty}^{\tau_{0}+i\infty}\frac{d\tau}{2\pi i} \int\limits
_{a(0)=
a(\tau)=a}{}D[a]
		e^{-I(\{a\})-\tau F}.
\end{equation}

	The common saddle point of the two remaining integrals corresponds to the
	energy condition found above for the Born-Oppenheimer equation, with the
	addition that the expectation value of the energy is found to be
	precisely the thermal expectation value
	\begin{equation}
\left\langle\hat{h}_{M}\right\rangle=\frac{\int{}{}D[\phi]exp[-I_{M}]\hat{h}_{M}}{\int{}{}D[\phi]
exp[-I_{M}]}=\frac{const}{\tau^{4}}.
\end{equation}
The energy condition thus is that found by the Born-Oppenheimer treatment above
\begin{equation}
\dot{a}^{2}+a^{2}-\frac{\Lambda}{3}a^{4}=-
\frac{16\pi}{3}\left\langle\hat{h}_{M}\right\rangle\equiv -\epsilon(\tau).
\end{equation}

Integrating this equation inside the tunneling domain,
 we can solve the equation by seeking for the self consistent solution of
\begin{equation}
\tau(a)=2\int_{a_{-}}^{a}\frac{da}{\sqrt{-\epsilon(\tau)+
a^{2}-\frac{\Lambda}{3}a^{4}}}.
\end{equation}
When $a=a_{+}$  we will identify this as the reciprocal temperature of
the physical domain.   That is
\begin{equation}
	\beta=2\int_{a_{-}}^{a_{+}}\frac{da}{\sqrt{-\epsilon(\beta)+a^{2}-
	\frac{\Lambda}{3}a^{4}}}.
\end{equation}
Furthermore we can then identify the quantity $W(a_{+})$ with the physical
entropy.
Thus we obtain the equation
\begin{equation}
	W(a_{+})= exp[S].
\end{equation}
For $a>a_{+}$,  $W(a)$ is independent of $a$, but maintains the weight
function of the statistical distribution.

If we were to continue the
path to $a$'s on the decreasing stage of the universe, since our paths all
begin at some $a$ go out to infinity retrace the path back to $a_{-}$,  we
will find an entropy which again increases when we penetrate the
tunneling region on recollapse of the universe.
Thus we find that the universe is ``born'' , beginning its temporal
expansion in a thermal inflationary state with a finite temperature as
measure in conformal time. This asymptotically de Sitter state has
necessarily a finite density of thermal bosons -like the stable black
hole case of radiation in a radiation cavity-.  We have elsewhere shown
that the Gibbons-Hawking\cite{GH} thermal de Sitter solution is
unstable.\cite{HW2}  The
crucial point of our argument is that though the formal procedure of
Hawking and coworkers found the suggestive result in which the black hole
and de Sitter cosmology appear thermal,  their approach misses the fact
that the presence of thermal matter is not only a reflection of the black
hole emission, but is fundamental to the explanation and the source of the
entropy of the problem. The attempts to show that the entropy is either
an entanglement entropy or an effect near the horizon are at least in that
respect in accord with our point of view. Thus the recent results of
finding due to these terms a black hole entropy differing from A/4 are
precisely in the spirit of our results for the cosmological case.
 But as to the source of the entropy we do not agree with these other points of
view.
 At least for the cosmology case we have shown explicitly the source of the
entropy to
be a mixing of matter states with the semi-classical gravity
, with no connection to particle states, no
connection to a separation of states inside and outside the horizon.

\section{Concluding remarks}
\abz We have demonstrated how both time in cosmology and the thermal state of
a cosmological asymptotically de Sitter universe arise from semi-classical
tunneling of gravity in the presence of quantum matter. Both of these
quantities have their origin in the mixing of semi-classical gravity and
matter in the tunneling region. The same mixing effect
provides the multiplicity of high energy states necessary to yield the
classical behavior of the gravity. We accept the point of view that the
problem is solved in terms of the specifics of ``our universe'', i.e.
that we require a positive cosmological constant in the early universe to
make it begin to evolve.  We have presented solutions only for a uniform,
isotropic universe.  Whether our results are limited to that case is not
clear, but I think it would appear that we at least require the universe
to begin with conformal time-like symmetry.  As we have stated above we
claim that as long as a single universe is the case in point and that we
are dealing with the wave function of the total universe, hence a closed
system time can only appear as intrinsic  quantity and only when gravity
is semi-classical.

	Left unresolved in the present work is the quesion to what extent the thermal
inflationary universe evolving with an intrinsic time given by the modified
Schroeding
equation is or is not in an essentially pure state.  This requires a detailed
analysis of the $\phi(\vec{x})$ integrals at both turning points. A
determination that
the integrals are sharply peeked at one value would establish the equivalence.
At the
lower turning point this is necessry to identify the density matrix with the
square of
the wave function.   At the upper turning point this is necessary to establish
that
the integration over the  $\phi(\vec{x})$ does not convert the square of the
wavefunction
into a density matrix. The argument of order interchange as a separate issue
is evidently
necessary for the result to be valid. The question of coherence for our results
thus cannot even be addressed until the above verification is accomplished.

Two major generalizations are necessary before these results can reach
their full impact.   In the first place we need to consider in detail a more
general model
of the matter which can both provide us with the switching cosmological
constant and the phase transition between inflation and Friedmann
cosmology. We have work in this direction in progress. Secondly we must
discover
 how to extend the
approach to black holes.  There may be some differences in even important
details but the basic physical source of the entropy would be expected to
be the same.\vfill

\vspace{1 em}
\noindent\large\bf Acknowledgments

\normalsize\rm
We would like to thank R. Brout, J. Bekenstein, C.J. Isham, J.Katz, O. Fonarev,
C. Kiefer
 O. Lechtenfeld, R.  Parentani and N.Dragon for useful conversations.
 A somewhat controversial discussion with
W. Unruh was the push for some of the point of view which crystallized in the
present
work.

\end{document}